\documentclass[conference]{IEEEtran}
\IEEEoverridecommandlockouts
\usepackage{cite}
\usepackage{amsmath,amssymb,amsfonts}
\usepackage[ruled,vlined,linesnumbered]{algorithm2e}
\usepackage{graphicx}
\usepackage{textcomp}
\usepackage{xcolor}
\usepackage{multirow}
\usepackage{booktabs}
\usepackage{array}
\usepackage{tikz}
\usetikzlibrary{arrows.meta,positioning,calc,fit,backgrounds}
\usepackage[hyphens]{url}
\usepackage{fancyhdr}
\usepackage{hyperref}
\usetikzlibrary{arrows.meta,positioning,calc,decorations.pathreplacing,fit,backgrounds}
\def\BibTeX{{\rm B\kern-.05em{\sc i\kern-.025em b}\kern-.08em
    T\kern-.1667em\lower.7ex\hbox{E}\kern-.125emX}}

\begin{document}

\title{Compiler Framework for 3D Neutral-Atom Quantum Computers}

\author{
\IEEEauthorblockN{
\begin{tabular}{c}
Chen Huang$^{1,2}$, Zhemin Zhang$^{2}$, Zhao Zhang$^{3}$, Xudong Lv$^{4,*}$, and
Zhiding Liang$^{1,5,*}$
\end{tabular}
}
\vspace{1ex}

\IEEEauthorblockA{\small%
\begin{tabular}{@{}c@{}}
$^{1}$Department of Computer Science and Engineering,\\ The Chinese University of Hong Kong, Hong Kong SAR, China\\
$^{2}$Open Quantum Intelligence\\
$^{3}$Max-Planck-Institut f\"ur Quantenoptik, 85748 Garching, Germany\\
$^{4}$SIOM, Chinese Academy of Sciences, Shanghai 201800, China\\
$^{5}$State Key Laboratory of Quantum Information Technologies and Materials,\\
The Chinese University of Hong Kong, Hong Kong SAR, China\\
\end{tabular}\\[3pt]
\footnotesize $^{*}$Corresponding authors: xudonglv@siom.ac.cn, zliang@cse.cuhk.edu.hk
}
}

\maketitle

\begin{abstract}
Neutral-atom quantum computers can now arrange atoms in three-dimensional tweezer arrays, yet every existing compiler assumes a flat geometry. We present Piqasso, a compiler that exploits the vertical axis by stacking storage, entanglement, and readout into distinct layers. Its pipeline pairs an analytical placement respecting axial-clearance optics with a router that brings gate partners together via short vertical hops—bypassing in-plane crossing conflicts through out-of-plane detours—and a multi-AOD scheduler that parallelizes transport across focal planes. On 34 circuits, Piqasso reduces atom transport distance by 2.1$\times$ over a state-of-the-art planar compiler, yielding up to 7.3$\times$ faster execution, 2.2× higher movement fidelity, and 1.8$\times$ fewer serialized transport rounds, with all gains widening at scale.
\end{abstract}

\begin{figure*}[t]
    \centering
    \includegraphics[width=\textwidth]{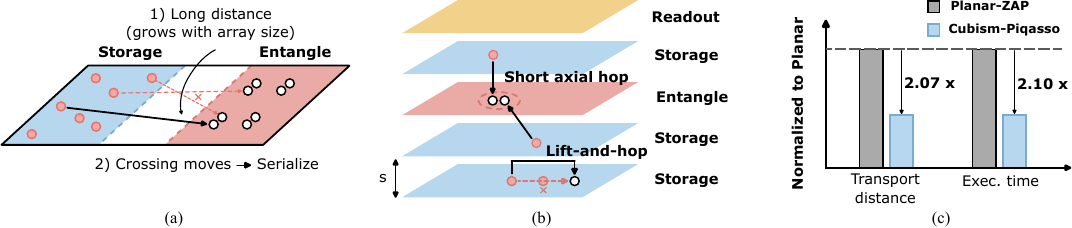}
    \caption{Why 3D compilation. (a)~A zoned planar architecture requires lateral transport that grows with array size, and the no-crossing constraint serializes conflicting moves. (b)~Piqasso stacks storage, an entanglement layer $z_E$, and a readout layer along $z$: gate pairs rendezvous via short axial hops, and blocked moves detour through a free transit layer (lift-and-hop). (c)~Across 34 circuits from QASMBench and VeriQBench, Piqasso moves atoms $2.1\times$ less and is faster on 32 of 34 circuits.}
    \vspace{-2ex}
    \label{fig:teaser}
\end{figure*}

\section{Introduction}
Neutral-atom quantum computers offer large qubit arrays, long coherence times~\cite{manetsch6100qubits2025}, and high-fidelity Rydberg interactions~\cite{levineParallelImplementationHighfidelity2019,everedHighfidelityParallelEntangling2023}. Recent experiments further demonstrate dynamic atom rearrangement~\cite{bluvsteinCoherentTransport2022,bluvsteinLogicalQuantumProcessor2024} and three-dimensional (3D) optical tweezer control~\cite{barredo3D2018,picard3DAOD2025,aol3DShuttle2025}. Because two-qubit gates require physical proximity, atom transport dominates execution on large arrays~\cite{tanOLSQDPQA2024,wangAtomique2024}. In a planar array, this transport is costly for three reasons. First, the distance atoms must travel to meet for a gate grows with array size, increasing execution time at scale. Second, atoms dephase during transport and while idle between moves, so longer travel directly reduces fidelity, with the loss compounding as circuits grow. Third, AOD-based moves cannot cross, so conflicting moves must be serialized, limiting transport parallelism. A third dimension addresses all three problems: gate pairs meet via a short axial hop whose length is independent of array size; shorter transport reduces dephasing; and blocked moves detour out of plane instead of waiting in line.

Atom transport is driven by AODs: a crossed-AOD frame holds all its atoms in one focal plane, whose tones may not cross during a move, and concurrent frames must not collide. Optical addressing adds an axial constraint: focused beams illuminate atoms above and below their target, so atoms in the same column must maintain a minimum axial clearance. These constraints make 3D compilation a new problem: joint placement, routing, and scheduling under coupled AOD and optical requirements.

We name this 3D layered architecture \emph{Cubism} and present \emph{Piqasso}, the compiler that targets it. The names nod to Cubism, the art movement that reassembled objects across multiple planes at once, and to Picasso, its foremost painter. Piqasso has three components: (i)~a clearance-aware initial placement that minimizes transport while respecting axial clearance; (ii)~conflict-aware non-planar routing through a lift-and-hop relaxation, where atoms detour over a free transit layer to bypass the no-crossing constraint and rendezvous on a dedicated entanglement layer with blockade-aware batching; and (iii)~a multi-AOD scheduler that legalizes simultaneous moves into non-crossing frames and parallelizes them across AOD subsystems. Figure~\ref{fig:teaser} contrasts this axial-rendezvous design with the zoned planar alternative.

We evaluate Piqasso on parameterized circuits and an application suite of 34 circuits from QASMBench~\cite{qasmbench} and VeriQBench~\cite{chen2022veriqbenchbenchmarkmultipletypes}, against a state-of-the-art (SOTA) zoned planar baseline (Planar-ZAP~\cite{huang2026}). By replacing the growing lateral transport of the zoned design with a short axial hop, Piqasso reduces total transport distance by $2.1\times$ on average and achieves faster execution on 32 of 34 circuits, with both advantages widening at scale.

In summary, this paper makes the following contributions:
\begin{itemize}
    \item We propose \emph{Cubism}, a 3D layered neutral-atom architecture that stacks storage, entanglement, and readout layers along $z$, motivated by the layer selectivity of 3D beam optics, and identify the optical constraints any compiler targeting it must satisfy.

    \item We present \emph{Piqasso}, the first end-to-end 3D neutral-atom compiler, with clearance-aware placement, lift-and-hop routing, and multi-AOD scheduling under realistic optical constraints.
    
    \item We quantify the cost and benefit of the third dimension across parameterized and application benchmarks. Against the Planar-ZAP baseline, axial rendezvous cuts transport distance by $2.2$--$2.7\times$ at the largest parameterized sizes and reduces execution time by up to $3.1\times$ on 100-qubit QFT, with the gap widening at scale. The shorter transport yields higher movement fidelity at the scales where the planar baseline degrades most. The 3D register also incurs ${\sim}3\times$ fewer serialization conflicts and requires $1.8\times$ fewer AOD transport rounds than ZAP.
\end{itemize}

The rest of this paper is organized as follows. \S\ref{sec:background} reviews neutral-atom hardware and the 3D optics underlying AOD/AOL control. \S\ref{sec:problem} formalizes the Cubism architecture and its movement, no-crossing, and axial-clearance constraints, and states the 3D compilation problem. \S\ref{sec:workflow} presents the Piqasso pipeline: clearance-aware placement, lift-and-hop routing, and multi-AOD scheduling. \S\ref{sec:setup} and \S\ref{sec:results} describe the experimental setup and results. \S\ref{sec:related} discusses related work, and \S\ref{sec:conclusion} concludes.
    
\section{Background}
\label{sec:background}

\subsection{Neutral Atom Quantum Computing}

Neutral-atom platforms encode qubits in atoms held by optical tweezers, arranged into programmable arrays by SLMs or AODs~\cite{barredo2016,endres2016}. Qubits are stored in two long-lived hyperfine ground states~\cite{saffmanQuantumInformationRydberg2010}. The platform exposes exactly two native gate types. Single-qubit gates use a two-tier drive: a global Raman carrier (or microwave field) drives $X_{\pm\pi/2}$ pulses on all atoms, while focused AC-Stark beams add site-resolved $R_z$ phases; any rotation decomposes into this basis~\cite{levineParallelImplementationHighfidelity2019,bluvsteinCoherentTransport2022,grahamMultiqubitEntanglementAlgorithms2022}. On unaddressed qubits the two carrier pulses cancel, making selection purely local: the focused light is diagonal in the computational basis, so axial leakage onto column neighbors causes only small, calibratable phase shifts rather than bit-flip errors.

The two-qubit gate uses the Rydberg blockade: two atoms within the blockade radius share a single Rydberg excitation, realizing a controlled-phase (CZ) gate~\cite{jakschFastQuantumGates2000,saffmanQuantumInformationRydberg2010,levineParallelImplementationHighfidelity2019}. CZ gates now execute in parallel at fidelities above $99.5\%$~\cite{everedHighfidelityParallelEntangling2023,bluvsteinLogicalQuantumProcessor2024}. Arbitrary single-qubit rotations together with CZ form a universal set~\cite{barenco1995}; we therefore assume the input has already been synthesized into this basis~\cite{qiskit2024}. The problem this paper addresses is bringing each CZ pair into blockade range via physical transport.

\subsection{3D Beam Optics}
\label{background-beam}

Every field that acts on the register (traps, gate light, readout light) is delivered as laser light. In a stacked register, these beams fall into two geometric classes: focused Gaussian beams, which propagate along the stacking ($z$) axis and address individual sites, and light sheets, which propagate in plane and illuminate an entire layer at once.

\begin{figure}[t]
    \centering
    \includegraphics[width=\linewidth]{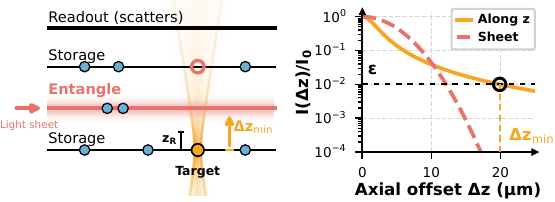}
    \caption{Propagation direction governs layer selectivity. Left: stacked register, side view. The workhorse beams (traps, tweezers, focused $R_z$ addressing) propagate along $z$ and thread the entire stack: a beam focused on its target atom (waist $w_0$) still illuminates the column neighbor through the slow on-axis tail, forcing the axial clearance $\Delta z_{\min}$. The in-plane Rydberg light sheet is confined to the entanglement layer $z_E$, and readout occurs only on the far-separated readout layer. Right: the two decay scales behind this asymmetry. Along $z$ the intensity falls only algebraically, reaching the crosstalk budget $\epsilon$ at $\Delta z_{\min}$, whereas the sheet's transverse Gaussian tail is exponentially suppressed well before. }
    \vspace{-2ex}
    \label{fig:beam}
\end{figure}

\textbf{Focused Gaussian beams.} As shown in Figure~\ref{fig:beam} (left, orange), a beam of waist $w_0$ propagating along the optical ($z$) axis broadens transversely as
\begin{equation}\label{eq:waist}
    w(z) = w_0\sqrt{1+\left(z/z_R\right)^2},
\end{equation}
where the Rayleigh length $z_R = \pi w_0^2/\lambda$ (with $\lambda$ the wavelength) sets the focal depth: beyond $|z|\sim z_R$ the spot diverges and both trap depth and addressing intensity fall off. The full 3D intensity of a beam centered on an addressed atom is
\begin{equation}\label{eq:gaussian}
    I(r,z)=\frac{I_0}{1+(z/z_R)^2}\,\exp\!\left(-\frac{2r^2}{w(z)^2}\right),
\end{equation}
where $r=\sqrt{x^2+y^2}$ is the transverse offset from the beam axis and $I_0$ the peak on-axis intensity. The two factors decay on very different scales: laterally the Gaussian suppresses illumination exponentially, whereas on axis ($r=0$) the intensity falls only algebraically as $I_0/[1+(z/z_R)^2]$. Since these beams propagate along $z$, neighboring layers sit on this slow axial tail: every such beam threads the entire stack.

This class includes the register's primary fields: the static SLM trap array, the mobile AOD/AOL tweezers, and the focused single-qubit addressing spots. Two constraints on compilation follow directly. First, axial confinement is much weaker than transverse (the trap is elongated over $\sim z_R$), so Piqasso bounds axial transport by a separate, lower speed $v_z<v_{xy}$. Second, a beam addressing one atom still illuminates atoms sharing its $(x,y)$ column at nearby heights, so atoms in a column must keep a minimum vertical separation $\Delta z_{\min}$; we call this the \emph{axial-clearance} constraint, and it couples placement across layers.

\textbf{Light sheets.} Rydberg excitation light can be shaped into a stripe that propagates in plane, near-uniform across the interaction region and Gaussian in thickness~\cite{bluvsteinLogicalQuantumProcessor2024}. For such a sheet the stacking direction is its transverse direction, so its $z$ profile follows the fast exponential tail (the ``sheet'' curve in Figure~\ref{fig:beam}, right). This sharp confinement is what makes a dedicated entanglement layer physically meaningful: every non-participating atom need only maintain an axial separation of at least $\Delta z_{E,\min}$ from the entanglement layer, a bound set jointly by residual excitation light and the blockade radius $r_b$. Zone-selective Raman beams admit the same light-sheet delivery.

The remaining fields occupy one of two selectivity extremes: the global single-qubit drive (a microwave field, or a Raman beam expanded to a waist of hundreds of microns whose Rayleigh range is meters) is homogeneous over the whole volume, so in-place single-qubit gates cost the same on every layer; conversely, resonant readout light scatters indiscriminately, which is why measurement is confined to a far-separated readout layer. Table~\ref{tab:beams} summarizes this taxonomy.

\begin{table*}[t]
    \caption{Delivery direction, spatial extent, and resulting layer ($z$) selectivity of the optical and microwave fields in a stacked register. Values are representative of current Rb/Cs tweezer platforms.}
    \label{tab:beams}
    \centering
    {\footnotesize
    \renewcommand{\arraystretch}{1.25}
    \begin{tabular}{@{}>{\raggedright\arraybackslash}m{3.6cm}>{\raggedright\arraybackslash}m{3cm}>{\raggedright\arraybackslash}m{6.7cm}>{\raggedright\arraybackslash}m{3.4cm}@{}}
    \toprule
    \textbf{Field} & \textbf{Delivery direction} & \textbf{Spatial profile / extent} & \textbf{Layer ($z$) selectivity} \\
    \midrule
    Static traps (SLM) & along $z$ & focused spot: $w_0\approx0.8$--$1.2~\mu$m, $z_R\approx2$--$5~\mu$m~\cite{barredo3D2018,levineParallelImplementationHighfidelity2019} & poor: algebraic \\
    \hline
    Mobile traps (AOD+AOL) & along $z$ & same spot~\cite{bluvsteinCoherentTransport2022} & poor: algebraic; beam crosses every layer while moving \\
    \hline
    Local 1Q addressing ($R_z$ only) & along $z$ & AC-Stark spot per site: $w_0\approx1$--$1.5~\mu$m~\cite{grahamMultiqubitEntanglementAlgorithms2022,bluvsteinCoherentTransport2022} & poor: algebraic \\
    \hline
    Plane 1Q drive & in plane & light sheet over one layer, thickness $w_z\approx5$--$10~\mu$m; or microwave $+$ per-layer light shift~\cite{kusano2025} & sharp: Gaussian; or frequency-selective \\
    \hline
    Global 1Q drive (carrier) & any & microwave: $\lambda\sim$cm~\cite{grahamMultiqubitEntanglementAlgorithms2022}; or expanded co-propagating Raman: $w_0\gtrsim10^2~\mu$m~\cite{bluvsteinCoherentTransport2022} & none, by design: uniform over the volume \\
    \hline
    Rydberg 2Q excitation & in plane ($420+1013$~nm, counter-prop.)~\cite{levineParallelImplementationHighfidelity2019} & flat-top stripe over the entanglement layer $z_E$~\cite{everedHighfidelityParallelEntangling2023,bluvsteinLogicalQuantumProcessor2024}, thickness $w_z\sim10~\mu$m & sharp: Gaussian; safe beyond $\Delta z_{E,\min}$ ($\ge r_b$) \\
    \hline
    Readout / imaging & onto readout layer~\cite{bluvsteinLogicalQuantumProcessor2024} & resonant light; scatters in all directions & none: isolate on far-separated readout layer \\
    \bottomrule
    \end{tabular}}
    \vspace{-2ex}
\end{table*}

\subsection{AOD-Based 3D Atom Transport}
\label{background-aod}

AODs steer laser beams by acoustic frequency modulation, repositioning traps continuously rather than in discrete hops~\cite{bluvsteinCoherentTransport2022}. In Cubism, a mobile AOD frame is generated by a crossed pair of AODs together with a double-pass acousto-optic lens (AOL) that sets a single focal height for every atom the frame carries~\cite{aol3DShuttle2025}. All atoms in one frame therefore share a common focal plane; reaching different target layers in parallel requires distinct frames.

Within a single frame, the column and row tones must remain strictly monotonic throughout the move: two atoms sharing a column (or row) translate in tandem, and their relative order can never invert. This no-crossing constraint~\cite{tanOLSQDPQA2024}, which must hold at every intermediate time, not only at the endpoints, is the fundamental limitation on which moves can share a frame; moves whose source-to-target ordering would cross must be assigned to separate frames.

The third dimension relaxes this limitation in two ways. First, with $n_{\mathrm{aod}}$ independent AOD subsystems, up to $n_{\mathrm{aod}}$ frames can run concurrently in one round, provided their spatial occupancies do not collide: two frames may share a focal plane as long as their in-plane footprints remain laterally separated. Second, and more importantly, a frame whose in-plane motion would cross an obstacle can instead lift to a free transit layer, translate there unobstructed, and lower back: a ``lift-and-hop'' maneuver with no planar analogue. Lift-and-hop converts the in-plane no-crossing problem into the simpler task of finding a free transit layer plus two short axial hops.

A transit layer requires no extra hardware: an atom in flight is held by the AOD/AOL tweezer alone, so it is simply a reserved band of focal heights free of static traps. The hop height is continuous, but the lift cannot be marginal: the moving tweezer is a $z$-propagating light column whose on-axis intensity decays only algebraically (Eq.~\eqref{eq:gaussian}), so every bypassed atom must stay outside the tweezer's axial crosstalk range, i.e., at least $\Delta z_{\min}$ from its focus. An atom threading between two occupied layers must clear both, which is why reserving dedicated transit layers is convenient, though not strictly required.

These constraints are why a 2D compiler cannot simply be extended to 3D: a planar router has no notion of focal planes, models a single control plane rather than potentially colliding frames, and never reasons about axial clearance.

\section{System \& Problem Formulation}
\label{sec:problem}

\subsection{The Cubism Architecture}

Cubism pairs a fixed SLM scaffold with mobile AOD/AOL transport: the SLM provides a multilayer register of $K$ layers at heights $z_1<\dots<z_K$, separated by at least $\Delta z_{\min}$ and spanning at most the AOL range $Z_{\max}$, while mobile AOD frames reconfigure atoms between these static sites. We take the layers uniformly spaced by $s$, so the fixed axial budget $Z_{\max}$ couples layer count and spacing through $(K-1)\,s\le Z_{\max}$. Rather than partitioning the plane laterally into storage and entanglement zones as in 2D zoned compilers, Cubism stacks its regions along $z$:
\begin{itemize}
    \item a \emph{storage region} occupying every layer except the entanglement and readout layers, where atoms are parked and single-qubit gates are applied in place;
    \item a dedicated \emph{entanglement layer} at height $z_E$ near the middle of the stack, where all two-qubit gates rendezvous for a global Rydberg pulse;
    \item a \emph{readout layer} at the top of the stack, to which completed qubits are migrated for measurement isolation.
\end{itemize}

Because rendezvous sites on $z_E$ are positioned continuously by AOD control rather than fixed to grid edges, each gate admits multiple feasible rendezvous configurations, giving the router additional freedom.

\subsection{Movement and Axial Clearance Constraints}

A \emph{frame} carries a set of column tones, a set of row tones, and one AOL focal height; each atom in the frame sits at the intersection of one column tone and one row tone in the frame's shared focal plane. Throughout a move, column tones must preserve their relative order, and likewise row tones: the orderings may never invert, even transiently. Two simultaneous moves may share a frame only if they share a focal plane and are mutually non-crossing; otherwise they must split into separate frames.

Simultaneous moves are greedily legalized into a near-minimal number of legal frames, each driven by its own AOD subsystem. All frames of a round execute concurrently provided no two collide, i.e., occupy focal planes within $\Delta z_{\min}$ of each other with overlapping in-plane footprints. A frame that would collide on its natural path can detour over a free transit layer (lift-and-hop); the rest serialize, so the transport-round count measures residual serialization.

A separate constraint arises from the addressing optics rather than AOD tone ordering. Because addressing beams have finite axial extent, atoms sharing the same $(x,y)$ coordinate must maintain sufficient vertical separation whenever one of them is addressed:
\begin{equation}\label{eq:axial-constraint}
    (x_i, y_i) = (x_j, y_j)
        \;\Longrightarrow\;
        |z_i - z_j| \ge \Delta z_{\min},
        \quad \forall i \neq j.
\end{equation}

This requirement is spatio-temporal: it must hold during initial placement and dynamically throughout movement and gate execution. It also limits vertical packing density, since an operation on one layer can temporarily block atoms elsewhere in the same column.

\subsection{Atom Movement Model}

Prior compilers assume uniform-acceleration trajectories with an isotropic Euclidean cost. We instead treat in-plane and axial transport as physically distinct, sequential channels. In-plane motion is driven by chirping the AOD tones; if the tweezer moves too fast, the finite trap depth cannot supply the required acceleration and the atom is lost. The in-plane peak speed is therefore capped at $v_{xy}<v_{\mathrm{loss}}$, with a safety margin below the empirical atom-loss threshold $v_{\mathrm{loss}}$~\cite{bluvsteinCoherentTransport2022}. Axial motion is driven by AOL focus tuning and, because axial confinement is weaker, uses a lower bound $v_z<v_{xy}$. A cross-layer move consists of an axial hop followed by an in-plane translation; the two durations add.

Trajectories follow a smooth minimum-jerk profile~\cite{flash1985}, $x(t)=d\,(6\sigma^5-15\sigma^4+10\sigma^3)$ with $\sigma=t/\Delta t$, ensuring atoms start and end at rest. This profile is smoother than the cubic ramps used experimentally~\cite{bluvsteinCoherentTransport2022}. We cap the peak rather than mean speed at the hardware limit $v_{xy}$, making our duration estimates conservative. The peak velocity of this profile is $15/8$ times the mean velocity $d/\Delta t$; capping it at the physical bounds gives a move duration of
\begin{equation}
    \Delta t = \frac{15}{8}\left(\frac{|d_z|}{v_z}+\frac{d_{xy}}{v_{xy}}\right),
\end{equation}
which reduces to a single term for purely in-plane or purely axial moves.

\subsection{Problem Formulation}

We formulate 3D neutral-atom compilation as a fidelity-driven optimization problem: given a quantum circuit and an initial atom configuration, the compiler determines atom placements, movement trajectories, and execution schedules to maximize overall circuit fidelity. Circuit execution time is not an independent objective but the compiler's primary lever for improving fidelity: all live qubits accumulate idle dephasing over the full schedule, and longer transport adds motional heating. Shortening the schedule therefore improves fidelity for every qubit simultaneously.

The overall execution fidelity is modeled as the product of operation-, transport-, and idle-related factors:
\begin{equation}
\label{eq:fidelity}
\begin{aligned}
    F=&\;\underbrace{(F_1)^{N_{g_1}}}_{\text{1Q gate}} \cdot
    \underbrace{(F_g^{\,2})^{N_{\mathrm{sp}}}}_{\text{carrier bkg.}} \cdot
    \underbrace{(F_2)^{N_{g_2}}}_{\text{2Q gate}} \cdot
    \underbrace{(F_h)^{N_h}}_{\text{Handover}}
    \cdot \underbrace{\textstyle\prod_{e}\bigl(1-\epsilon_e^{\text{xt}}\bigr)}_{\text{Crosstalk}} \\ &\; \times
    \underbrace{\textstyle\prod_{m}\bigl(1-\epsilon_m^{\text{heat}}\bigr)}_{\text{Transport heating}} \cdot
    \underbrace{\textstyle\prod_{q\in \mathcal{Q}}\exp\left(-t_\text{idle}^q/T_2^*\right)}_{\text{Decoherence}},
\end{aligned}
\end{equation}
where $F_1$ is the composite per-target single-qubit fidelity (two carrier pulses plus the local $R_z$). The carrier-background factor accounts for the volume-uniform drive: $F_g$ is the fidelity of one global carrier pulse ($99.97\%$--$99.99\%$ in Raman randomized benchmarking~\cite{bluvsteinCoherentTransport2022,manetsch6100qubits2025}), and $N_{\mathrm{sp}}=\sum_{s\in\mathcal{S}_1}\bigl(|\mathcal{Q}^{\mathrm{live}}_s|-|\mathcal{Q}^{\mathrm{tgt}}_s|\bigr)$ counts spectator exposures, since every not-yet-measured qubit absorbs the two carrier pulses of each single-qubit stage $s\in\mathcal{S}_1$ even when it is not a target. $F_2$ denotes the two-qubit gate fidelity, $F_h$ the fidelity of each \textsc{pick}/\textsc{drop} handover between an SLM trap and an AOD frame, $\epsilon_m^{\text{heat}}$ the motional-heating infidelity of move $m$, and $t_\text{idle}^q$ the total time qubit $q$ spends idle (waiting in a trap while not being gated or transported) until it is migrated to the readout layer or the circuit ends. Dephasing at the trap coherence time $T_2^*$ accrues even while an atom is parked.

The crosstalk term is a product over crosstalk-exposure events, where $\epsilon_e^{\text{xt}}$ is the infidelity of event $e$. In a 3D register, crosstalk can arise from several sources: stray addressing light leaking across focal planes, residual Rydberg tails, and spectator atoms falling within an active beam waist. Rather than modeling each source individually, Piqasso confines every interaction to reserved safe positions: the axial-clearance constraint (Eq.~\eqref{eq:axial-constraint}) together with the dedicated entanglement and transit layers keeps all non-participating atoms outside any crosstalk-exposure region. Under this design the crosstalk factor reduces to unity; we retain the term so that a relaxed layout violating these safe positions would be penalized accordingly.

The gate counts $N_{g_1}$ and $N_{g_2}$ are fixed by the input circuit, so the gate-fidelity factors in Eq.~\eqref{eq:fidelity} are beyond the compiler's control. The compiler can only reduce the remaining losses: handovers, transport heating, and idle dephasing. Per-qubit idle times are known only after a complete schedule exists, but all shrink together when the schedule gets shorter. The compiler therefore uses total circuit execution time as its practical optimization target: subject to the clearance and speed constraints above, every placement, routing, and round-packing decision scores candidates by transport distance or time, keeping moves short and thereby limiting both heating and idle exposure.

\section{Piqasso Compilation Workflow}\label{sec:workflow}

\begin{figure*}[t]
    \centering
    \includegraphics[width=\linewidth]{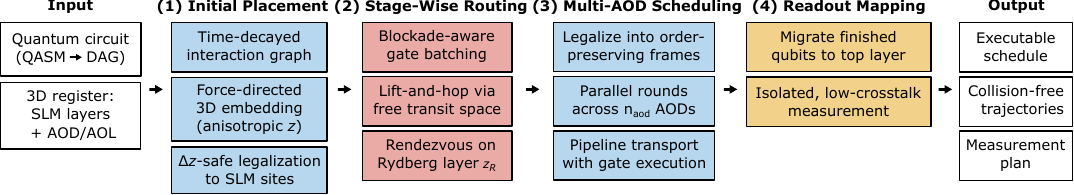}
    \caption{The Piqasso compilation pipeline: (i)~clearance-aware initial placement, (ii)~stage-wise routing via entanglement-layer rendezvous, (iii)~multi-AOD transport-round scheduling with cross-round pipelining, and (iv)~readout mapping to the readout layer; the output is an executable schedule with collision-free trajectories and a measurement plan.}
    \vspace{-2ex}
    \label{fig:pipeline}
\end{figure*}

Figure~\ref{fig:pipeline} illustrates the Piqasso workflow. Given an input quantum circuit and an initial atom configuration, the compiler produces a physically executable schedule by jointly optimizing qubit placement, atom routing, and operation timing under AOD and optical constraints.

The pipeline has four stages. (i)~Clearance-aware initial placement pulls frequently interacting qubits closer while enforcing the axial clearance. (ii)~Stage-wise routing brings each two-qubit gate to a rendezvous on $z_E$ under the crossed-AOD constraints. (iii)~Multi-AOD scheduling packs non-colliding frames into concurrent transport rounds and pipelines the descent of one pulse round with the ascent of the next. (iv)~Readout mapping migrates completed qubits to the readout layer, isolating measurement from ongoing computation.

Throughout, we use four scheduling units. A \emph{stage} groups logically independent gates of the circuit; a \emph{pulse round} groups gates that share one global Rydberg pulse; a \emph{frame} is the set of moves carried by one AOD subsystem; and a \emph{transport round} groups frames that execute concurrently.

\subsection{Analytical Initial Placement}
\label{subsec:initial-placement}

Rather than solving a discrete combinatorial placement problem, whether by simulated annealing~\cite{tanEnola2025,lin2025} or greedy heuristics~\cite{wangAtomique2024,huang2026}, we decompose placement into a continuous optimization followed by hardware-aware legalization, following the analytical VLSI placement tradition~\cite{lu2015eplace}. The rationale is that circuit interaction structure induces spatial locality: frequently interacting qubits, especially in early stages, belong close together. The 3D hardware adds geometric and optical constraints, particularly along $z$, that are naturally handled in a separate legalization step.

Given a stage-ordered circuit with two-qubit gate sets $\{\mathcal{G}_0, \mathcal{G}_1, \dots\}$, we construct a weighted interaction graph $G = (\mathcal{Q}, E, W)$, where each vertex represents a qubit and edge weight $w_{ij}$ captures the interaction intensity between qubits $q_i$ and $q_j$. For each two-qubit gate $(q_i, q_j)$ at stage $t$, we accumulate a temporally decayed contribution:
\begin{equation}
    w_{ij} \leftarrow w_{ij} + e^{-\delta t},
\end{equation}
where $\delta$~$> 0$ prioritizes early-stage interactions.

We assign each qubit $q_i$ a continuous coordinate $\mathbf{p}_i \in \mathbb{R}^3$ and minimize the surrogate energy:
\begin{equation}
\begin{aligned}
E(\mathbf{P}) = &\;
\sum_{i<j} w_{ij}\bigl(\|\mathbf{p}_i-\mathbf{p}_j\|_{s} - d^\star\bigr)^2 \\
&\; + \lambda_{\mathrm{rep}} \sum_{i<j} \phi\!\left(\|\mathbf{p}_i-\mathbf{p}_j\|_{s}\right),
\end{aligned}
\label{eq:energy}
\end{equation}
where $d^\star$ is the preferred interaction distance, $\lambda_{\mathrm{rep}}>0$ weights the repulsion, and $\phi(d)=1/d$ is a short-range Coulomb-like penalty that grows as two qubits approach: its gradient yields an inverse-square repulsive force that keeps distinct qubits from collapsing onto a common site. Here $\|\cdot\|_{s}$ denotes the anisotropic distance
\begin{equation}
\|\mathbf{p}_i-\mathbf{p}_j\|_{s} = \sqrt{\Delta x_{ij}^2 + \Delta y_{ij}^2 + s_z^2\,\Delta z_{ij}^2},
\label{eq:aniso-norm}
\end{equation}
where the scalar $s_z$ weights axial ($z$) separations relative to in-plane ones; its value is fixed by the transport cost structure below.

This weighting is counterintuitive at first. Since axial transport is slower ($v_z<v_{xy}$), one might expect $s_z>1$ to penalize $z$-separations. The rendezvous architecture inverts this expectation. Every two-qubit gate meets at $z_E$, so each atom pays only its own short axial hop regardless of which storage layer it occupies; the inter-qubit vertical separation contributes almost nothing to transport cost. What matters is the in-plane offset. Setting $s_z>1$ would save no axial travel while collapsing the placement into a single plane and forfeiting the register's axial capacity. Setting $s_z<1$ instead spreads qubits across layers, relieving the in-plane crowding that actually drives transport. Candidate sites span all static SLM sites on the storage layers; the readout layer is reserved for measurement and excluded from initial placement.

We minimize Eq.~\eqref{eq:energy} using an iterative force-directed procedure~\cite{fruchterman1991}. At each iteration, the net force on qubit $q_i$ is 
\begin{equation}
    \mathbf{F}_i = - \nabla_{\mathbf{p}_i} E(\mathbf{P}),
\end{equation}
which combines an attractive spring term that pulls strongly interacting qubits toward their preferred separation $d^\star$, with a short-range repulsion that prevents overlap. The position is then updated as
\begin{equation}
\mathbf{p}_i \leftarrow \mathbf{p}_i + \eta_k \mathbf{F}_i,
\end{equation}
where $\eta_k$ is a decreasing step size. All positions are clipped to the valid 3D bounding region.

Since atoms must occupy discrete SLM sites, we legalize by greedily mapping each qubit to a physical site, processing qubits in descending order of weighted degree $d_i = \sum_j w_{ij}$ and choosing for each the site $s \in \mathcal{S}$ minimizing
\begin{equation}
    C_{\mathrm{legal}}(q_i, s) = \|\mathbf{p}_i - s\|^2 + \lambda_z \cdot \mathbb{I}_{\mathrm{violate}}(s),
\label{eq:legal}
\end{equation}
where $\mathbb{I}_{\mathrm{violate}}$ indicates violation of the axial-clearance constraint and $\lambda_z$ is a large penalty weight. The constraint is soft: a violating site is chosen only when no compliant site is available, which keeps the legalizer applicable to arbitrary site sets. In the evaluated register the constraint additionally holds by construction, since the SLM scaffold spaces its layers at $\geq\Delta z_{\min}$ (Table~\ref{tab:parameters}) and transport moves atoms between such sites via the reserved entanglement and transit layers.

\subsection{Stage Routing via Entanglement-Layer Rendezvous}
\label{subsec:stage-placement}

After initial placement, the compiler processes the circuit stage by stage. Single-qubit gates are applied in place, since they are individually addressed and require no transport. Two-qubit gates execute through the rendezvous mechanism. The goals are to (i)~assign each gate a rendezvous pair satisfying the gate distance $d_g$, (ii)~batch as many gates as possible under one pulse without blockade cross-coupling, and (iii)~realize the required transport with non-crossing, lift-and-hop AOD moves.

Because positions on $z_E$ are continuously addressable by the AOD, we place each gate's rendezvous near the in-plane midpoint of its two atoms to minimize transport. For a gate $(q_i,q_j)$ with current in-plane center $(c_x,c_y)$, the two atoms are assigned the symmetric sites
\begin{align}
    & p_0 = \bigl(c_x - \tfrac{d_g}{2},\; c_y,\; z_E\bigr),\\
    & p_1 = \bigl(c_x + \tfrac{d_g}{2},\; c_y,\; z_E\bigr),
\end{align}
so that $\|p_0 - p_1\| = d_g$. This keeps each pair compact at the entanglement layer while leaving the rendezvous center free to track the atoms' storage locations.

Under a single global Rydberg pulse, two gates cross-couple if any atom of one pair falls within the blockade radius $r_b$ of the other pair. To guarantee independence, two gates may share a pulse only if their rendezvous centers are separated by at least $r_b + d_g$:
\begin{equation}
    \left|\mathbf{c}_a - \mathbf{c}_b\right| \ge r_b + d_g
    \ \Rightarrow\ 
    \text{gates } a,b \text{ may share a pulse.}
\label{eq:blockade-sep}
\end{equation}
The stage's two-qubit gates are partitioned into pulse rounds by a greedy independent-set construction on the resulting blockade-conflict graph: each gate joins the first pulse round whose members all satisfy Eq.~\eqref{eq:blockade-sep}, otherwise a new pulse round is opened. This rendezvous packing replaces the in-plane interaction-zone packing of 2D zoned compilers.

Because a single global Rydberg beam illuminates $z_E$, pulses fire sequentially, one per pulse round. The transport between pulses, however, is pipelined: the descent of pulse round $k$'s atoms back to storage and the ascent of pulse round $k{+}1$'s atoms up to $z_E$ involve disjoint atoms on complementary paths, so both merge into one move batch. Each batch is greedily legalized into a near-minimal number of non-crossing, single-plane AOD frames, using lift-and-hop to a free transit layer whenever this removes an in-plane crossing more cheaply than opening a new transport round. Since pulse round $k{+}1$ fires only after the whole batch completes, no atom is illuminated at $z_E$ while still in flight. Algorithm~\ref{alg:stage-placement} summarizes the complete procedure.

\begin{algorithm}[t]
\caption{Stage Routing via Entanglement-Layer Rendezvous}
\label{alg:stage-placement}
\KwIn{Current mapping $\mathcal{M}$, stage two-qubit gates $\mathcal{G}^{(2)}$, entanglement layer $z_E$, blockade radius $r_b$, gate distance $d_g$}
\KwOut{Updated mapping $\mathcal{M}'$ and executed pulse rounds}
$\mathcal{M}' \leftarrow \mathcal{M}$;\quad $\mathcal{M}_{\mathrm{store}} \leftarrow \mathcal{M}$ \tcp*{storage sites to return to}
\ForEach{gate $(q_i,q_j)\in \mathcal{G}^{(2)}$}{
    $\mathbf{c} \leftarrow$ midpoint of $\mathcal{M}'[q_i]$, $\mathcal{M}'[q_j]$ in $xy$\;
    assign rendezvous sites $(p_0,p_1)$ at $z_E$ around $\mathbf{c}$ at separation $d_g$\;
}
// Greedy independent set on blockade-conflict graph (Eq.~\eqref{eq:blockade-sep})\;
partition $\mathcal{G}^{(2)}$ into pulse rounds $\{\mathcal{R}_1,\mathcal{R}_2,\dots\}$\;
\ForEach{pulse round $\mathcal{R}_k$}{
    $\mathcal{T} \leftarrow \mathcal{M}'$ with $q_i \!\to\! p_0$, $q_j \!\to\! p_1$ for each $(q_i,q_j)\in\mathcal{R}_k$\newline
    \phantom{$\mathcal{T} \leftarrow$} and each $\mathcal{R}_{k-1}$ atom returned to $\mathcal{M}_{\mathrm{store}}$ \tcp*{pipelined descent}
    $\textsc{RouteBatch}(\mathcal{M}' \!\to\! \mathcal{T})$ \tcp*{non-crossing frames + lift-and-hop}
    $\mathcal{M}' \leftarrow \mathcal{T}$\;
    fire one global Rydberg pulse on all pairs in $\mathcal{R}_k$\;
}
$\mathcal{B} \leftarrow \mathcal{M}'$ with each atom of the last round returned to $\mathcal{M}_{\mathrm{store}}$\;
$\textsc{RouteBatch}(\mathcal{M}' \!\to\! \mathcal{B})$;\quad $\mathcal{M}' \leftarrow \mathcal{B}$\;
\Return{$\mathcal{M}'$}\;
\end{algorithm}

\subsection{Stage Scheduling and Multi-AOD Transport Rounds}
\label{subsec:scheduling}

\begin{figure}[t]
    \centering
    \includegraphics[width=1\linewidth]{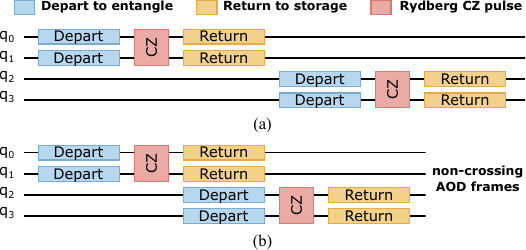}
    \caption{Cross-round pipelining. (a)~In a planar machine the return of pulse round and the ascent of pulse round move in opposite directions through a shared plane, forcing tone crossings that serialize the handoff. (b)~In Cubism these moves are near-axial and pack into non-crossing frames; lift-and-hop detours any residual crossing through a free transit layer, so the axial round trips do not inflate the transport-round count.}
    \vspace{-2ex}
    \label{fig:scheduling}
\end{figure}

Piqasso schedules the logical circuit into execution stages with an ASAP (As-Soon-As-Possible) policy: each gate is placed in the earliest stage in which both qubits are free, so independent gates naturally group into the same stage for parallel execution. This stage list defines the dependency order that the router respects.

Within a stage, execution time is dominated by transport to and from the entanglement layer, not by gate pulses. The third dimension shortens this transport, but every gate now requires a vertical round trip. Cross-round pipelining and lift-and-hop mitigate this overhead (Figure~\ref{fig:scheduling}).

Each move batch is packed into transport rounds of mutually non-colliding frames, one AOD subsystem per frame, with no cap on concurrent frames. A frame that collides on its natural path is admitted via lift-and-hop whenever the marginal time it adds is below the cost of a new transport round. The batch's execution time is the transport-round count times the slowest-atom time per round. This produces a finer-grained schedule than a global barrier $t_{\text{gate}} \geq \max_i(t_{\text{arrival}}^{(i)})$; we report the transport-round count against a no-lift baseline to measure residual serialization directly.
 
\subsection{Readout Mapping}

At the end of each stage, the router evacuates atoms in two groups: atoms whose last gate has completed are migrated to the top readout layer for measurement isolation, while atoms still needed by later stages return to the nearest free storage-layer site. Both evacuation batches reuse the same non-crossing, lift-and-hop transport.

Measurement isolation improves both fidelity and parallelism. Because the readout layer is axially separated from the computation layers, the $500~\mu$s resonant imaging~\cite{bluvsteinLogicalQuantumProcessor2024} of an evacuated qubit overlaps with ongoing computation on the other layers; the axial gap, rather than a time barrier, protects live qubits from scattered photons. Our cost model reflects this: a measured qubit is charged only for the transport to the readout layer, never for the readout dwell. Once resident there, it stops accumulating idle coherence loss and no longer counts as a spectator for global carrier pulses. Mid-circuit measurements~\cite{norcia2023} therefore overlap with computation rather than extending circuit execution time.

\section{Evaluation Setup}
\label{sec:setup}
\subsection{Baseline}

We compare against ZAP~\cite{huang2026}, a SOTA open-source compiler for the zoned planar architecture, referred to as \emph{Planar-ZAP} throughout. In this design, storage and entanglement occupy separate lateral zones in one plane, so every two-qubit gate requires lateral transport whose distance grows with the array size. For a fair comparison, Planar-ZAP uses the same hardware parameters as Piqasso (Table~\ref{tab:parameters}), so the results isolate the architectural difference under identical hardware assumptions.

\subsection{Hardware Parameters and Compiler Configuration}

Table~\ref{tab:parameters} summarizes the hardware model, drawn from SOTA experiments: gate fidelities and durations follow Evered et al.~\cite{everedHighfidelityParallelEntangling2023}, handovers ($15~\mu$s each) dominate per-move latency~\cite{bluvsteinCoherentTransport2022}, and $T_2^*=12.6$\,s, the hyperfine coherence time demonstrated on a $6{,}100$-qubit tweezer array~\cite{manetsch6100qubits2025}, means decoherence accumulates over total execution time rather than individual operations.

Transport is anisotropic: in-plane peak speed is capped at $v_{xy}=0.55~\mu\text{m}/\mu\text{s}$~\cite{bluvsteinCoherentTransport2022}, a safety margin below the atom-loss threshold $v_{\mathrm{loss}}$; axial motion uses the lower bound $v_z=0.40~\mu\text{m}/\mu\text{s}$; durations follow the minimum-jerk model with axial and transverse phases additive. Layers are spaced at $21~\mu\text{m}$ (${\geq}\Delta z_{\min}$) within the AOL range $Z_{\max}$; gate pairs rendezvous at separation $d_g=2.5~\mu\text{m}$ and share a pulse only beyond $r_b$.

\begin{table}[t]
    \centering
    \scriptsize
    \caption{Hardware model: operation fidelity/duration and 3D AOD/AOL transport parameters. Readout figures are reference values; the cost model charges readout only through the transport that migrates qubits to the readout layer.}
    \label{tab:parameters}
    \resizebox{\columnwidth}{!}{%
    \begin{tabular}{|l|c|c|c|c|}
    \hline
    \multirow{2}{*}{Operation} & \multicolumn{2}{c|}{Fidelity} & \multicolumn{2}{c|}{Duration} \\
    \cline{2-5}
     & Symbol & Value & Symbol & Value \\
    \hline
    1Q gate (target: $2\times$carrier$\,+\,$local $R_z$) & $F_1$ & $0.9975$~\cite{grahamMultiqubitEntanglementAlgorithms2022} & $t_1$ & $0.625~\mu$s \\
    Global carrier $\pi/2$ (per spectator) & $F_g$ & $0.9997$~\cite{bluvsteinCoherentTransport2022} & \multicolumn{2}{c|}{(within 1Q stage)} \\
    2Q gate & $F_2$ & $0.995$~\cite{everedHighfidelityParallelEntangling2023} & $t_2$ & $0.36~\mu$s~\cite{everedHighfidelityParallelEntangling2023} \\
    Handover (\textsc{pick}/\textsc{drop}) & $F_h$ & $0.999$~\cite{bluvsteinCoherentTransport2022} & $\tau_{\mathrm{ho}}$ & $15~\mu$s \\
    Readout & -- & $0.998$~\cite{bluvsteinLogicalQuantumProcessor2024} & -- & $500~\mu$s~\cite{bluvsteinLogicalQuantumProcessor2024} \\
    Coherence & \multicolumn{2}{c|}{--} & $T_2^*$ & $12.6~$s~\cite{manetsch6100qubits2025} \\
    \hline
    \multicolumn{3}{|l|}{Transport / geometry parameter} & Symbol & Value \\
    \hline
    \multicolumn{3}{|l|}{Transverse peak-speed} & $v_{xy}$ & $0.55~\mu$m$/\mu$s \\
    \multicolumn{3}{|l|}{Axial (AOL) peak-speed} & $v_z$ & $0.40~\mu$m$/\mu$s \\
    \multicolumn{3}{|l|}{Atom-loss speed threshold} & $v_{\mathrm{loss}}$ & $1.0~\mu$m$/\mu$s \\
    \multicolumn{3}{|l|}{Min layer separation} & $\Delta z_{\min}$ & $20~\mu$m \\
    \multicolumn{3}{|l|}{AOL axial range} & $Z_{\max}$ & $136~\mu$m \\
    \multicolumn{3}{|l|}{Rydberg blockade radius} & $r_b$ & $5~\mu$m \\
    \multicolumn{3}{|l|}{Intra-pair gate distance} & $d_g$ & $2.5~\mu$m \\
    \hline
    \end{tabular}%
    }
    \vspace{-2ex}
\end{table}

The Piqasso placement hyperparameters are fixed across all benchmarks and configurations. For the interaction graph, we set the temporal decay $\delta = 0.1$ and the preferred interaction distance $d^\star = 2.5$, matching the gate distance $d_g$. For the surrogate energy, we set the axial discount $s_z = 0.15$ and the repulsion weight $\lambda_{\mathrm{rep}} = 50$. The force-directed procedure runs for $200$ iterations with an initial step size of $2.0$ scaled by $0.95$ per iteration, and legalization uses the axial-violation penalty $\lambda_z = 10$.

\subsection{Benchmarks}

We evaluate on two tiers of benchmarks (Table~\ref{tab:benchmarks}). The first tier consists of three parameterized families swept in size to trace how the 3D advantage scales: QFT~\cite{coppersmith2002} (dense global interactions), QAOA Max-Cut on random graphs~\cite{farhi2014} (irregular parallel interactions), and Cuccaro ripple-carry adders~\cite{cuccaro2004} (sequential dependency chains). Fixed instances from these families also serve as testbeds for the conflict, ablation, and sensitivity studies.

The second tier is an application suite of $34$ fixed circuits from QASMBench~\cite{qasmbench} and VeriQBench~\cite{chen2022veriqbenchbenchmarkmultipletypes}, spanning nine functional classes and $7$--$118$ qubits. Every circuit compiles under all three evaluated configurations, so all baselines are compared on an identical set; suite instances that overlap with a swept family are independent library implementations. All circuits are transpiled to the $\{\text{CZ},\,\text{single-qubit}\}$ gate set~\cite{qiskit2024} and compiled at each circuit's compact footprint under the hardware model of Table~\ref{tab:parameters}.

\begin{table}[t]
\centering
\begin{scriptsize}
\caption{Benchmarks: three parameterized families swept for scaling analysis, and a $34$-circuit application suite in nine functional classes ($7$--$118$ qubits). $n$: qubit counts; CZ: two-qubit-gate count range after transpilation.}
\label{tab:benchmarks}
\resizebox{\columnwidth}{!}{%
\begin{tabular}{|l|l|r|}
\hline
\multicolumn{3}{|l|}{\emph{Parameterized families (swept for scaling)}} \\
\hline
Family & Sizes ($n$) & CZ \\
\hline
QFT & $10,18,29,50,63,100$ & $81$--$3481$ \\
QAOA MaxCut & $10,20,30,50$ & $16$--$342$ \\
Cuccaro adder & $4,10,64$ & $10$--$455$ \\
\hline
\multicolumn{3}{|l|}{\emph{Application suite}} \\
\hline
Class & Benchmarks ($n$) & CZ \\
\hline
Arithmetic & adder ($10,28,64,100,118$), & $34$--$2286$ \\
 & multiplier ($15,45$), square\_root ($18$) & \\
Fourier \& QPE & qft ($18,29,63$), qpe ($9$), pe ($101$) & $43$--$3549$ \\
Search \& lin.\ alg. & grover ($79$), hhl ($7$) & $86$--$912$ \\
Comm.\ \& state prep & ghz ($40,78$), bv ($100$) & $39$--$99$ \\
Ham.\ simulation & ising ($10,66,98$) & $90$--$194$ \\
Machine learning & qcnn ($16$), dnn ($51$), & $123$--$601$ \\
 & swap\_test ($83$), qugan ($111$) & \\
Variational & hf-vqe ($12$), qaoa ($30,64,100$) & $72$--$360$ \\
Random \& volume & qv ($32$), supremacy ($8{\times}8,10{\times}10$) & $205$--$1488$ \\
Error correction & seca ($11$), qec9xz ($17$) & $29$--$80$ \\
\hline
\end{tabular}%
}
\end{scriptsize}
\vspace{-2ex}
\end{table}

\section{Evaluation Results}
\label{sec:results}

Our evaluation answers three questions, each tied to a planar cost identified in the introduction. (i)~\emph{Time}: does the short axial hop cut transport distance and execution time, and does the advantage grow with circuit size (\S\ref{sec:eval-time})? (ii)~\emph{Fidelity}: does shorter transport improve execution fidelity at the scales where the planar baseline suffers most (\S\ref{sec:eval-fidelity})? (iii)~\emph{Parallelism}: does the third dimension ease the serialization imposed by the no-crossing constraint (\S\ref{sec:eval-parallel})? A final set of ablations (\S\ref{sec:eval-ablation}) traces the gains to individual components and shows that a modest AOD budget, already within reach of current hardware, suffices to realize them.

\subsection{Transport Distance and Circuit Execution Time}
\label{sec:eval-time}

\begin{figure*}[t]
    \centering
    \includegraphics[width=1\linewidth]{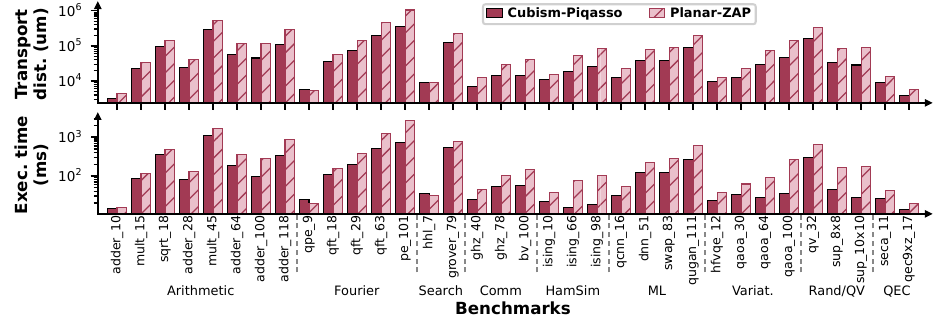}
    \caption{Per-circuit transport distance (top) and circuit execution time (bottom) across the full $34$-circuit application suite, for Cubism-Piqasso and the Planar-ZAP baseline (both log scale). Circuits are grouped by functional class and sorted by qubit count within each class. Piqasso moves atoms less on every circuit and runs faster on 32 of 34, with the gap widening on larger circuits; the two exceptions (hhl\_n7 and qpe\_n9) are small enough that the fixed axial hop is not yet amortized. }
    \vspace{-2ex}
    \label{fig:suite}
\end{figure*}

The central claim of the 3D design is that a gate needs only a short axial hop, not a lateral traversal that grows with the array. Figure~\ref{fig:suite} confirms this across all $34$ circuits. Piqasso moves atoms less on every circuit (top) and runs faster on all but two (bottom), with both gaps widening at scale. On average it cuts transport distance and execution time by $2.1\times$, and the speedup reaches $7.3\times$ on the largest circuits. The two exceptions, hhl\_n7 and qpe\_n9, are small enough that the fixed axial hop is not yet amortized, the crossover we examine next.

\begin{figure}[t]
    \centering
    \includegraphics[width=1\linewidth]{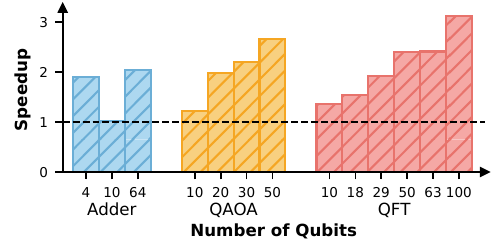}
    \caption{Execution time speedup of Piqasso over the Planar-ZAP baseline on the three parameterized families, as a function of circuit size. Within every family (Adder, QAOA, QFT) the speedup grows with qubit count.}
    \vspace{-2ex}
    \label{fig:zoned}
\end{figure}

The three parameterized families reveal how the advantage scales with qubit count (Figure~\ref{fig:zoned}). Below roughly ten qubits the two architectures perform similarly; above that crossover the speedup grows steadily, reaching $2.0$--$3.1\times$ at the largest sizes. The range reflects circuit structure: Adder, whose sequential carry chain offers little transport to parallelize, sits at the low end, while the transport-heavy QFT benefits most. In all cases the speedup follows directly from the reduction in transport distance.

\begin{figure*}[t]
    \centering
    \includegraphics[width=1\linewidth]{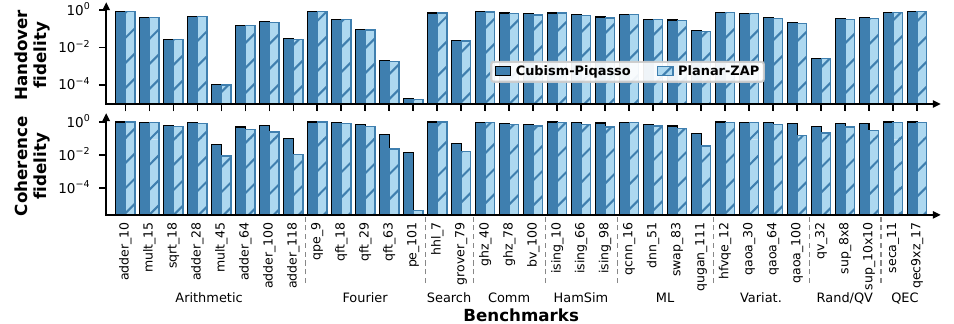}
    \caption{Execution fidelity on the application suite, per circuit, grouped by functional class. The two panels are the factors of movement fidelity $F_h\cdot F_{\mathrm{coh}}$ (log scale, higher is better) for the Piqasso vs. Planar-ZAP baseline. Top: the handover factor $F_h$, essentially identical across compilers because both perform a comparable number of trap-to-AOD transfers. Bottom: the coherence factor $F_{\mathrm{coh}}$, which drives the entire gap: the shorter axial transport leaves atoms less time to dephase at $T_2^*$. The gain is largest for the transport-bound classes and negligible for QEC, which barely moves atoms.}
    \label{fig:fidelity}
\end{figure*}

\subsection{Execution Fidelity}
\label{sec:eval-fidelity}

Transport carries its own fidelity cost: atoms dephase at $T_2^*$ while being moved or parked, and every \textsc{pick}/\textsc{drop} handover incurs a transfer error. The single- and two-qubit gate factors are fixed by the input circuit and cancel across compilers, so we isolate the compiler-controlled movement fidelity $F_h\cdot F_{\mathrm{coh}}$.

Figure~\ref{fig:fidelity} plots the two factors of movement fidelity per circuit, and they tell opposite stories. The handover factor $F_h$ (top) is essentially identical: the Piqasso and Planar-ZAP bars overlap on every circuit, because both pipelines perform a comparable number of trap-to-AOD transfers. If anything $F_h$ favors Piqasso slightly on the largest circuits, where the axial rendezvous needs marginally fewer handovers (geometric mean $0.142$ vs. $0.134$; the per-circuit ratio never drops below one), so handover error never works against the third dimension.

The gain lives entirely in coherence (bottom). The idle-plus-transport dephasing factor $F_{\mathrm{coh}}$ rises from a geometric mean of $0.26$ under Planar-ZAP to $0.53$ under Piqasso, because replacing the $O(\sqrt{n})$ lateral transport with a short axial hop cuts the time atoms spend in motion or parked between rounds. The combined movement fidelity $F_h\cdot F_{\mathrm{coh}}$ improves by a geometric-mean $2.2\times$, from a negligible $1.0\times$ on QEC, which barely transports atoms, to $8.6\times$ on the transport-bound Fourier class. The advantage grows with circuit size, where planar transport is most costly: on the ${\geq}64$-qubit circuits it reaches $4.1\times$ ($0.011\!\rightarrow\!0.047$). The third dimension thus converts shorter transport directly into preserved coherence.

\subsection{Transport Parallelism}
\label{sec:eval-parallel}

\begin{figure*}[t]
    \centering
    \includegraphics[width=1\linewidth]{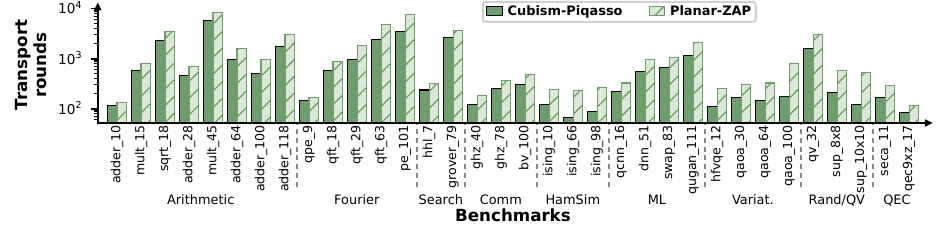}
    \caption{AOD transport rounds per circuit across the $34$-circuit application suite (log scale); fewer is better. The round count measures residual serialization: each round groups non-colliding AOD frames that execute concurrently. Piqasso issues fewer rounds on all $34$ circuits, because short axial hops generate fewer no-crossing conflicts than the lateral paths of the zoned baseline.}
    \vspace{-2ex}
    \label{fig:rounds}
\end{figure*}

The third planar cost is serialization: the no-crossing constraint forces conflicting moves into separate transport rounds, and a zoned register funnels all gate traffic through the same lateral paths. We quantify the relief using two measurements: the AOD transport-round count relative to Planar-ZAP and the number of conflicted frames each architecture generates.

The third dimension relieves serialization directly. Although every gate adds an axial ascent and descent, the Cubism register issues $1.8\times$ fewer AOD transport rounds than Planar-ZAP (Figure~\ref{fig:rounds}; geometric mean $0.54\times$, fewer rounds on all $34$ circuits, up to $4.6\times$ on qaoa\_n100; $2.2\times$ over the ${\geq}64$-qubit circuits). The extra focal planes absorb the added axial moves by running frames concurrently where the zoned register's shared lateral paths force serialization.

The transport-round count measures the effect; we also measure the cause. The compiler counts a frame as conflicted when its natural transport path overlaps the footprint of every round it could otherwise join on some focal plane, exactly the situation that the planar no-crossing constraint can resolve only by serialization. Each conflicted frame is then either lifted onto a free transit layer and merged into an existing round, or forced to open a new one.

On QFT-50, $938$ of $3{,}693$ frames ($25\%$) are conflicted under Piqasso; Planar-ZAP generates $2{,}822$ conflicted frames, $3.0\times$ more, because the shared entanglement zone funnels all traffic through the same lateral paths and has no third dimension to detour through. QAOA-30 shows the same pattern at lower contention: $90$ vs.\ $282$ conflicted frames ($3.1\times$). Conflicts are counted under each compiler's own move set, so the comparison measures how much serialization pressure each architecture generates. Short axial hops rarely cross, and this reduction in conflicts, rather than lift resolution, is the primary mechanism behind the transport-round reduction of Figure~\ref{fig:rounds}.

\begin{figure*}[t]
    \centering
    \includegraphics[width=\textwidth]{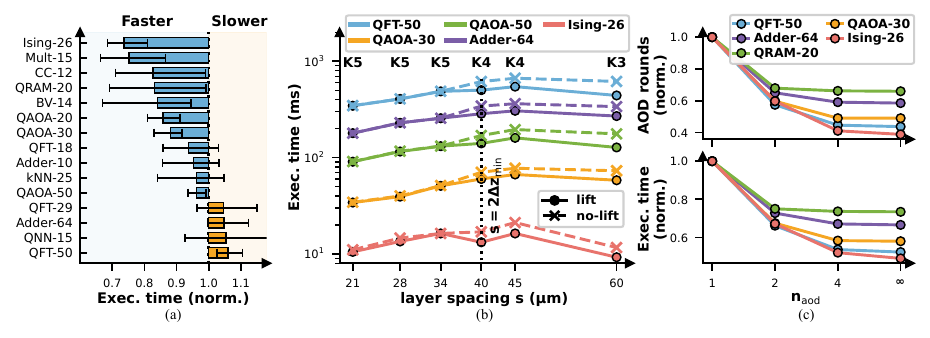}
    \caption{Ablation and sensitivity studies. (a)~Component ablation: per-circuit execution time under analytical placement normalized to a random-placement baseline. Random placement is averaged over five seeds; whiskers span the min--max ratio across seeds. 
    (b)~Layer-spacing co-design across five circuits shows the circuit execution time versus layer spacing $s$ with and without lift-and-hop. 
    (c)~Sensitivity to the AOD-subsystem budget $n_{\mathrm{aod}}$ for five circuits spanning dense (QFT-50), chain (Adder-64), tree (QRAM-20), irregular (QAOA-30), and local (Ising-26) structure, each normalized to its own $n_{\mathrm{aod}}=1$ point: concurrency reduces AOD transport rounds (top) and execution time (bottom) but is essentially saturated by $n_{\mathrm{aod}}{=}4$ on all five.}
    \label{fig:sensitivity}
\end{figure*}

\subsection{Ablation and Sensitivity Studies}
\label{sec:eval-ablation}

\textbf{Component ablation.}

We isolate the contribution of Piqasso's two core components by ablating each within the same 3D register ($5$ layers, unlimited AOD subsystems; Figure~\ref{fig:sensitivity}a). Replacing the clearance-aware analytical placement with an interaction-agnostic random baseline (averaged over five seeds) slows execution on $11$ of $15$ circuits, by up to $26\%$ (Ising) and $25\%$ (multiplier), where the interaction graph exposes spatial locality the placer can exploit. On all-to-all QFT and the $64$-qubit ripple-carry adder the two placements agree to within $6\%$: with every qubit pair interacting, or a single long carry chain, little locality remains to capture. Disabling lift-and-hop costs at most $1.0\%$ extra transport rounds and $0.6\%$ execution time at the dense main register, a finding the transit-layer trade-off below explains. The two components thus address complementary regimes.

\textbf{Lift-and-hop and the transit-layer trade-off.}
At the dense main register, lift-and-hop itself contributes little to the transport-round advantage. A transit layer must keep every bypassed atom at least $\Delta z_{\min}$ from the moving tweezer's focus, and the dense $21$-$\mu$m stack leaves no trap-free height between storage layers. The only candidates are the entanglement layer and the band above the readout layer, whose long axial hops rarely pass the cost-aware admission test.

This is a geometric claim about spacing, so we test it by sweeping $s$ directly. Because the axial budget $Z_{\max}$ is fixed, each step co-varies $s$ and $K$; we re-run five circuits (QFT-50, QAOA-30, QAOA-50, Adder-64, Ising-26) at each setting with and without lift-and-hop (Figure~\ref{fig:sensitivity}b). The effect appears exactly where the geometry predicts. For $s<2\Delta z_{\min}$ no trap-free transit layer fits between adjacent layers and lift-and-hop saves ${\le}2\%$ on the four larger registers (up to $8\%$ on the compact $26$-qubit Ising stack, whose sparse footprint occasionally leaves a trap-free detour). At $s\ge2\Delta z_{\min}$ a reserved transit layer fits inside every inter-layer gap, lift-and-hop resolves about half of all conflicted frames, and it recovers $14$--$23\%$ of execution time on the $K=4$ registers and $20$--$28\%$ on the sparse $K=3$ register.

The absolute optimum nonetheless remains the dense stack. Widening $s$ works against itself: it lengthens every axial hop to $z_E$ and, by thinning the register to fewer layers, packs more atoms onto each plane and increases in-plane transport. These two costs together outweigh the transit-layer savings. The main evaluation therefore adopts $s=21~\mu$m and treats lift-and-hop as the mechanism that keeps sparser registers, which hardware may prefer for optical robustness or addressing margin, competitive. This is a co-design result: layer spacing trades hop latency and in-plane crowding against routing freedom, and the compiler quantifies both sides.

\textbf{AOD-budget sensitivity.}
Multiple AOD subsystems let the 3D register realize its parallelism, and a budget already within reach of current hardware suffices. We sweep $n_{\mathrm{aod}}$ from $1$ to unlimited (Figure~\ref{fig:sensitivity}c) across five circuits of differing interaction structure: QFT-50 (dense), Adder-64 (chain), QRAM-20 (tree), QAOA-30 (irregular), and Ising-26 (local).

With a single AOD the axial ascent and descent of every gate serialize, forfeiting much of the register's round savings. Relaxing the budget lets non-colliding frames run concurrently, cutting AOD transport rounds by $56\%$ (QFT-50: $3{,}693\!\rightarrow\!1{,}613$) and $51\%$ (QAOA-30: $392\!\rightarrow\!192$) from $n_{\mathrm{aod}}{=}1$ to the unlimited limit, with corresponding $48\%$ and $42\%$ drops in execution time. Adder-64, QRAM-20, and Ising-26 cut rounds by $42\%$, $34\%$, and $61\%$, respectively.

The curves flatten quickly: at $n_{\mathrm{aod}}=4$ every circuit is within $6\%$ of its unlimited round count, and within $2\%$ for four of the five (QAOA-30 and QRAM-20 essentially exact), because concurrency saturates once it exceeds the number of focal planes with sufficient axial clearance. Four AOD subsystems already sit near the current hardware ceiling, so this saturation point is realizable rather than asymptotic: a near-term machine with only a few AOD subsystems realizes essentially all of the reported gains, and the unlimited-AOD assumption of the main evaluation is a convenience rather than a requirement.

\section{Related Work}
\label{sec:related}

Every neutral-atom compiler we are aware of assumes a strictly two-dimensional geometry~\cite{compilerSurvey2025}. These compilers follow one of two planar architectures. Monolithic DPQA compilers reconfigure a single global-Rydberg array: OLSQ-DPQA~\cite{tanOLSQDPQA2024} synthesizes layouts exactly via SMT but scales poorly, Enola~\cite{tanEnola2025} reaches thousands of qubits by serializing Rydberg stages under near-optimal edge coloring, Q-Pilot~\cite{wangQPilot2024} adds flying ancillas, and divide-and-shuttle~\cite{huang2025dasatomdivideandshuttleatomapproach} partitions circuits to cut movement. Within this line, Atomique~\cite{wangAtomique2024}, Parallax~\cite{parallax2024}, and PAC~\cite{chen2025} treat AOD control constraints as first-class, respectively driving multiple AOD arrays, sacrificing packing density for non-crossing, and compiling hardware regions in parallel. Zoned compilers (PowerMove~\cite{ruan2025}, ZAP~\cite{huang2026}, reuse-aware compilation~\cite{lin2025}, routing-aware placement~\cite{routingAwarePlacement2025}, Mantra~\cite{jang2025}, Search-Smarter~\cite{searchSmarter2026}) split the plane into storage and entanglement zones, so every interaction pays a lateral traversal. In both architectures, parallelism is bounded by in-plane non-crossing and Rydberg-stage serialization, with no axial dimension to relieve either.

A parallel thread treats qubit motion itself as a compilation target: QC-Daemon~\cite{qcDaemon2025} casts atom rearrangement as a reinforcement-learning game, trapped-ion QCCD compilers route ions through segments and junctions~\cite{murali2020,durandau2023}, and movable-logical-qubit lattice surgery~\cite{herzog2025} extends the idea to superconducting codes. Each confirms that scheduling movement is decisive, and each routes within a planar grid.

The hardware motivating an axial dimension now exists: 3D AODs and acousto-optic lenses move tweezers volumetrically~\cite{picard3DAOD2025,lu2026,aol3DShuttle2025}, and plane-selective addressing~\cite{kusano2025} over static multilayer scaffolds~\cite{schlosserTalbot2023,barredo3D2018} makes a layered register concrete. No existing compiler models the axial coordinate. Piqasso fills this gap by treating that coordinate as a schedulable resource: the no-crossing constraints that Parallax avoids by sacrificing density and Enola absorbs by serializing become layer-local and bypassable via lift-and-hop, a gate becomes a short axial hop rather than a lateral traversal, and the constraints unique to 3D (inter-frame plane collisions and the axial clearance $\Delta z_{\min}$) are handled jointly by placement, routing, and scheduling.

\section{Conclusion and Outlook}
\label{sec:conclusion}

We presented Piqasso, the first compiler that treats the axial coordinate of a neutral-atom processor as a first-class, schedulable resource. From the optics of crossed AODs and a double-pass AOL we derived the constraints governing parallel 3D transport (the one-plane-per-frame property, the no-crossing constraint, and the axial clearance $\Delta z_{\min}$) and addressed them with a unified pipeline: clearance-aware analytical placement, a lift-and-hop router, and a multi-AOD scheduler that pipelines transport across successive pulse rounds.

Stacking storage, an entanglement layer, and a readout layer along $z$ turns each two-qubit gate into a short axial rendezvous instead of lateral transport that grows as $O(\sqrt{n})$. On the $34$-circuit application suite, Piqasso moves atoms $2.1\times$ less than the Planar-ZAP baseline and, beyond ${\sim}10$ qubits, runs $2.1\times$ faster on average and up to $7.3\times$ on the largest circuits, with the advantage widening at scale. We position Piqasso as a reference baseline for future 3D placement, routing, and hardware co-design.

Several directions remain open. The register need not be a uniform cuboid: circuit-adapted layer shapes, or staggered packings in which adjacent planes share no axial column, correspond to different settings of $\Delta z_{\min}$ and the plane-collision test and thus fall within Piqasso's constraint model. These configurations invite hardware-compiler co-design of denser layouts together with the parameters $K$, $\Delta z_{\min}$, and $Z_{\max}$. The same lift-and-hop and non-crossing machinery also degrades gracefully to a single layer, so Piqasso can serve as the planar back end of a unified 2D/3D compiler.

\section*{Acknowledgment}
We thank Yuan Xu for the invaluable discussions.

\bibliographystyle{IEEEtranS}
\bibliography{refs}

\end{document}